\documentclass[twoside,12pt]{article}
\usepackage{epsfig}
\usepackage{amssymb}

\newcommand{\be}{\begin{equation}}
\newcommand{\ee}{\end{equation}}
\newcommand{\bea}{\begin{eqnarray}}
\newcommand{\eea}{\end{eqnarray}}

\begin{document}

\title{ \vspace{0.2cm} QCD Spin Physics: 
Partonic Spin Structure of the Nucleon\footnote{Talk presented by W.~Vogelsang}}
\author{D. de Florian,$^1$ R. Sassot,$^1$ M. Stratmann,$^2$ W.\ Vogelsang$^3$ \\
\\
$^1$Departamento de Fisica, Universidad de Buenos Aires,\\
Ciudad Universitaria, Pabellon 1 (1428) Buenos Aires, Argentina\\
$^2$Physics Department, Brookhaven National Laboratory,\\
Upton, NY 11973, U.S.A.\\
$^3$Institute for Theoretical Physics,
                Universit\"{a}t T\"{u}bingen,\\
                Auf der Morgenstelle 14,
                D-72076 T\"{u}bingen, Germany}
\maketitle
\begin{abstract} We discuss some recent developments concerning
the nucleon's helicity parton distribution functions: New preliminary data from jet
production at RHIC suggest for the first time a non-vanishing polarization 
of gluons in the nucleon. SIDIS measurements at COMPASS provide 
better constraints on the strange and light sea quark helicity distributions. 
Single-longitudinal
spin asymmetries in $W$-boson production have been observed at RHIC 
and will ultimately give new insights into the light quark and anti-quark 
helicity structure of the nucleon.
\end{abstract}
\section{Introduction}

QCD spin physics has been driven by
the hugely successful experimental program of polarized 
deeply-inelastic lepton-nucleon scattering (DIS)~\cite{leader}. 
One of the most important results has been the finding 
that the quark and anti-quark spins (summed over all flavors) provide only 
about a quarter of the nucleon's spin, $\Delta \Sigma\approx 0.25$ 
in the proton helicity sum rule~\cite{helsr}:
\begin{equation} \label{HSR}
\frac{1}{2}=\frac{1}{2}\Delta \Sigma + \Delta G+L_q+L_g \; . \label{ssr}
\end{equation} 
This result implies that sizable contributions to the nucleon spin should come 
from the gluon spin contribution $\Delta G$, or from 
orbital angular momenta $L_{q,g}$ of partons. 
To determine the other contributions to the nucleon spin 
has become a key focus of the field. In the present article, 
we describe some of the recent developments of the field.
We focus on current efforts to determine the helicity parton
distributions of the nucleon and on the latest experimental results.

The helicity structure of the nucleon is foremost described by
its twist-two helicity parton distribution functions, 
\begin{equation} \label{qdef}
\Delta f(x,Q^2)  \equiv f^+(x,Q^2) \; - \; 
f^-(x,Q^2)
\,\,\,\,\;\;\;\;\;\, 
(f=u,d,s,\bar{u},\bar{d},\bar{s},g)    \; ,
\end{equation}
$f^+$ ($f^-$)
denoting the number density of partons with the same (opposite) helicity 
as the nucleon's, as a function of momentum fraction $x$ and scale 
$Q$. QCD predicts the
$Q^2$-dependence of the densities through the spin-dependent
Dokshitzer-Gribov-Lipatov-Altarelli-Parisi  (DGLAP)
evolution equations~\cite{dglap1}:
\begin{equation} \label{dglapeq}
\frac{d}{d \ln Q^2} \left( \begin{array}{c}
\! \Delta q \!
\\ \! \Delta g \! \end{array} \right)(x,Q^2) 
= \left( \begin{array}{cc}
\! \Delta P_{qq}(\alpha_s,x) &  \Delta  P_{qg}(\alpha_s,x) \! \\
\! \Delta P_{gq}(\alpha_s,x) & \Delta P_{gg}(\alpha_s,x) \!
\end{array} \right) \;\otimes \;
\left( \begin{array}{c}
\!  \Delta q \! \\ \!  \Delta g \!
\end{array} \right) \left( x,Q^2 \right) \; ,
\end{equation}
where $\otimes$ denotes a convolution, and the splitting functions $\Delta P_{ij}$ 
are evaluated in QCD perturbation theory~\cite{dglap1,mvn,vm}.

The contributions $\Delta \Sigma(Q^2)$ and $\Delta G(Q^2)$
in the helicity sum rule~(\ref{HSR}) are given by
\begin{eqnarray} \label{sigma}
\Delta \Sigma(Q^2) &=& \int_0^1 \left( \Delta u+\Delta\bar{u}+
 \Delta d+\Delta\bar{d}+ \Delta s+\Delta\bar{s}\right)(x,Q^2) dx
\equiv \Delta \Sigma_u +  \Delta \Sigma_d +
\Delta \Sigma_s \;,\\[2mm]
\Delta G(Q^2)&=&\int_0^1 \Delta g(x,Q^2)dx\;.
\end{eqnarray}
$\Delta \Sigma$ is independent of $Q^2$ at the lowest order. 
The distributions have a proper field-theoretic definition. For example, 
in case of $\Delta g$ it is given by~\cite{mano}
\begin{equation}
\Delta g (x,Q^2) =
\frac{i}{4\pi\,x\, P^+} \int d\lambda \, {\rm e}^{i\lambda x P^+}\,
\langle P,S| G^{+\nu}(0)\, \tilde{G}^+_{\;\;\nu} 
(\lambda n)|P,S\rangle\Big|_{Q^2} \;, \label{opme}
\end{equation}
written in $A^+=0$ gauge. Here, $G^{\mu\nu}$ is the QCD field strength tensor, 
and $\tilde{G}^{\mu\nu}$ its dual. The integral of $\Delta g(x,Q^2)$ 
over all momentum fractions $x$ becomes a local operator only in $A^+=0$ 
gauge and then coincides with $\Delta G(Q^2)$~\cite{helsr}. 

\begin{figure}[b]

\vspace*{-0.6cm}
\hspace*{1cm}
\epsfig{figure=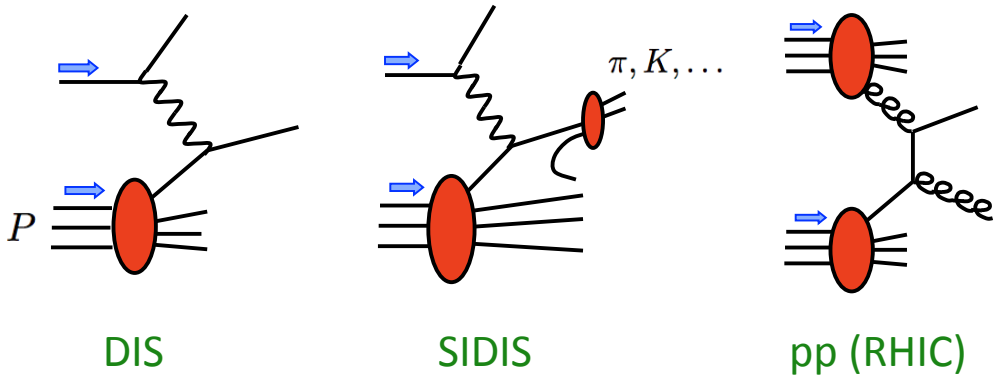,width=1.1\textwidth}

\vspace*{-8.6cm}
\caption{\label{probes} Parton-model Feynman diagrams for the processes
constraining nucleon helicity structure.}
\vspace*{0.cm}
\end{figure}
The helicity parton distributions may be probed in spin asymmetries
for reactions at large momentum transfer. The probes used so far
are inclusive and semi-inclusive deep-inelastic lepton scattering 
(DIS and SIDIS, respectively), and $pp$ scattering at large transverse
momentum, see Fig.~\ref{probes}. Polarized DIS and SIDIS experiments have 
been carried out at SLAC, CERN, DESY and the Jefferson Laboratory \cite{leader}
and mostly constrain the quark and anti-quark helicity distributions. 
RHIC at BNL~\cite{rhicrev,rhicrev1} is the first polarized proton-proton
collider, operating at $\sqrt{s}=200$ and 500~GeV. The measurement of gluon polarization 
in the proton is a major focus and strength of RHIC. 

The basic theoretical concept that underlies much of 
spin physics is the factorization theorem. It states that large
momentum-transfer reactions may be factorized into
long-distance pieces that contain the desired information on the
spin structure of the nucleon in terms of its {\em universal} parton
densities, and parts that are
short-distance and describe the hard interactions of the
partons. The latter can be evaluated using perturbation theory, 
thanks to the asymptotic freedom of QCD.
As an example, we consider the reaction $pp\to \pi X$, 
where the pion is produced at high transverse 
momentum $p_T$, ensuring large momentum transfer. 
The statement of the factorization theorem~\cite{collinsfact} is then:
\begin{eqnarray}
\label{eq:eq2}
d\Delta \sigma &=&\sum_{a,b,c}\, 
\Delta f_a \,\otimes \,\Delta f_b  \,\otimes\,
d\Delta \hat{\sigma}_{ab}^{c} \,\otimes \,D_c^{\pi} 
\end{eqnarray}
for the polarized cross section, where $\otimes$ denotes a convolution. 
The $D_c^{\pi}$ are the pion fragmentation functions. The sum in Eq.~(\ref{eq:eq2}) 
is over all  contributing partonic channels $a+b\to c + X$, with
$d\Delta \hat{\sigma}_{ab}^{c}$ the associated spin-dependent partonic cross
section. Factorization is valid up to corrections that are suppressed as 
inverse powers of the hard scale. In general, a leading-order estimate of (\ref{eq:eq2})
merely captures the main features, but does not usually provide a 
quantitative understanding. 
Only with knowledge of the next-to-leading order (NLO)
QCD corrections to the $d\Delta \hat{\sigma}_{ab}^{c}$
can one reliably extract information on the parton distribution functions 
from the reaction. By now, NLO corrections are available for most
of the processes relevant in polarized high-energy scattering~\cite{nlocomp}.

Independent information on the nucleon's helicity distributions
may be obtained by using SU(2) and SU(3) flavor symmetries. The integrals of the
flavor non-singlet combinations turn out to be proportional
to the nucleon matrix elements of the quark 
non-singlet axial currents, $\langle P,S \,|\, \bar{q} \,
\gamma^{\mu}\, \gamma^5 \,\lambda_i \,q \,|\, P,S \rangle$. 
Such currents typically occur in weak interactions, and 
one may relate the matrix elements to the $\beta$-decay 
parameters $F,D$ of the baryon octet. One finds
\begin{eqnarray} \label{a3a8s}
&&\Delta \Sigma_u - \Delta \Sigma_d = F+D=1.267 ,\nonumber \\
&&\Delta \Sigma_u + \Delta \Sigma_d -2 \Delta\Sigma_s=3F-D\approx 0.58 \ .
\end{eqnarray}
If valid, the second relation when combined with Eq.~(\ref{sigma}) gives
that $\Delta \Sigma=0.58 +3 \Delta\Sigma_s$, so that a small 
quark spin contribution to the proton spin implies a large negative
strange quark contribution. Fairly significant violations of SU(3) symmetry
have been predicted based on heavy baryon chiral perturbation theory~\cite{savage}. 
Lattice investigations of this issue have begun but are not yet conclusive~\cite{bali}.

\section{Nucleon helicity structure: status 2009}

\begin{figure}[t]

\vspace*{-0.3cm}
\hspace*{-0.5cm}
\epsfig{figure=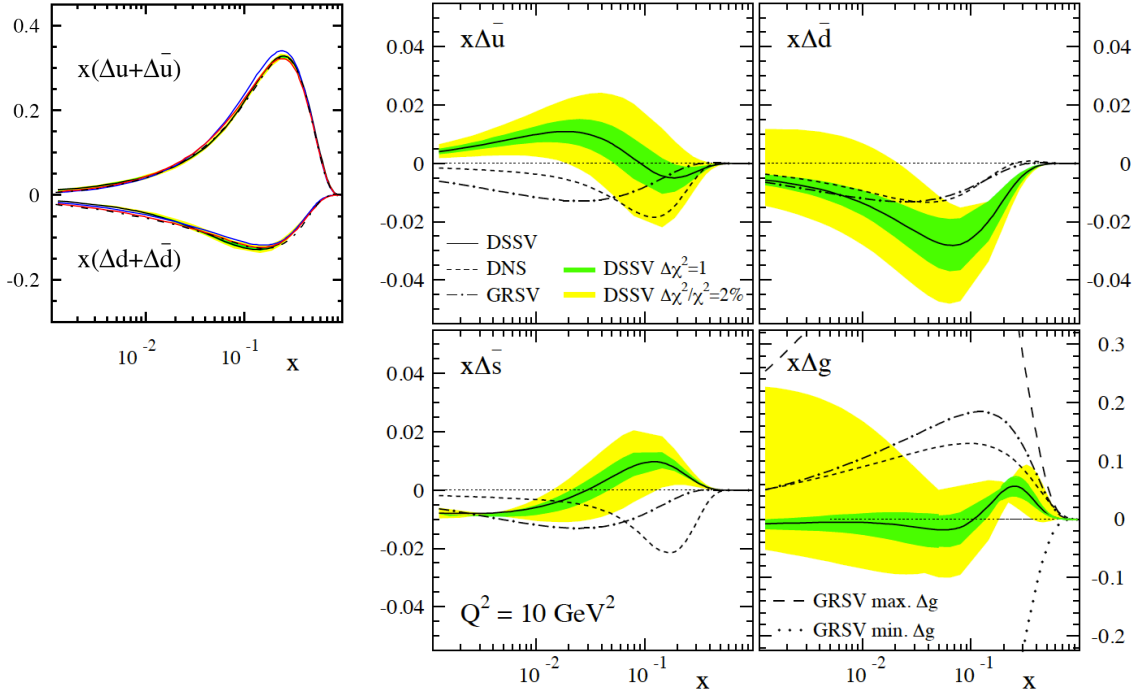,width=1.1\textwidth}

\vspace*{-4.5cm}
\caption{\label{pdfstatus} Present status of the nucleon's NLO helicity distributions
according to the global ana\-ly\-sis of Ref.~\cite{dssv}. The solid center lines
show the best-fit result. The shaded bands provide uncertainty estimates, using a criterion of 
$\Delta\chi^2 = 1$ (inner bands) or $\Delta\chi^2/\chi^2 = 2\%$ (outer bands)
as allowed tolerance on the $\chi^2$ value of the fit. Also shown are results from
earlier analyses~\cite{grsv,dns} of nucleon spin structure from lepton scattering data
alone.}
\vspace*{0.cm}
\end{figure}

In recent publications~\cite{dssv}, we have presented the first
next-to-leading order (NLO)
``global'' QCD analysis of the nucleon's helicity distribution from DIS,
semi-inclusive DIS (SIDIS), and $pp$ scattering at RHIC.  
We have used a Mellin moment method for the analysis. 
Our results are shown in Fig.~\ref{pdfstatus}, along with estimates 
of their uncertainties. The shaded bands in Fig.~\ref{pdfstatus} 
show the distributions that are allowed if one permits an overall
increase of $\Delta\chi^2 = 1$ (green) or 
$\Delta\chi^2/\chi^2 = 2\%$ (yellow). As one can see,
the ``total'' $\Delta u+\Delta\bar{u}$ and $\Delta d+\Delta\bar{d}$ helicity 
distributions are very well constrained. This is expected since these
distributions are primarily determined by the large body
of inclusive DIS data. Our results agree well with the distributions obtained in previous
and other recent analyses~\cite{grsv,dns,lss,bb,bbs} which considered only the
lepton scattering data. 

The sea anti-quark distributions still carry rather large 
uncertainties, even though they are better constrained now
than in previous analyses, thanks to the advent of more precise 
SIDIS data and of a new set of fragmentation functions~\cite{DSS} 
that describes the observables well in the unpolarized case. 
We find that the sea appears not to be SU(2)-flavor 
symmetric: the $\Delta \bar{u}$ distribution is mainly positive, while the 
$\Delta \bar{d}$ anti-quarks carry opposite polarization. This pattern has
been predicted at least qualitatively by a number of 
models~\cite{grsv,models}. 
Already based on the Pauli principle one would expect that 
if valence-$u$ quarks primarily spin along the proton spin direction, 
$u\bar{u}$ pairs in the sea will tend to have the $u$ quark polarized 
opposite to the proton. Hence, if such pairs are in a spin singlet, 
one expects $\Delta\bar{u} > 0$ and, by the same reasoning, $\Delta\bar{d} 
< 0$. We note that the uncertainties in SIDIS are still quite large, and it 
is in particular difficult to quantify the systematic uncertainty of the results
related to the fragmentation mechanism at the relatively modest 
energies available. 

The strange sea quark density shows a sign change. At moderately
large $x\sim 0.1$, it is constrained by the SIDIS data, which prefer a positive $\Delta s$.
On the other hand, the inclusive DIS data combined with the 
constraints from baryon $\beta$-decays demand a negative integral of $\Delta s$.
As a consequence, $\Delta s$ obtains its negative integral purely from the contribution
from low-$x$. Interestingly, there are initial lattice determinations of
the integral $\Delta \Sigma_s$~\cite{bali}, which point to small values.
It is clearly important to understand the strange contribution
to nucleon spin structure better.

Constraints on the spin-dependent gluon distribution $\Delta g$ predominantly come 
from RHIC. As can be seen from Fig.~\ref{pdfstatus}, the gluon distribution turns out to 
be small in the region of momentum fraction, $0.05\lesssim x\lesssim 0.2$, 
accessible at RHIC, quite possibly having a node. At $Q^2=10$~GeV$^2$, the integral 
over the mostly probed $x$-region is found to be almost zero, $\int_{0.05}^{0.2} dx 
\Delta g(x)=0.005\pm 0.06$, where the error is obtained for a variation of $\chi^2$ by
one unit. Thus, on the basis of~\cite{dssv}, 
there are no indications of a sizable contribution of gluon
spins to the proton spin. We also note that 
a way to access $\Delta g$ in lepton-nucleon scattering at HERMES and 
COMPASS is to measure final states that select the photon-gluon fusion process,
heavy-flavor production and high-$p_T$ hadron or hadron-pair
production~\cite{compass,hermesdg}. 
These data were not included in the analysis~\cite{dssv}, mostly
because of the fact that success of the perturbative-QCD hard-scattering 
description had not been established for these observables in the kinematic 
regime of interest here. At least for single-inclusive high-$p_T$ hadrons 
in $\gamma p\to h^{\pm} X$ it has now been found that
QCD hard-scattering does appear to be applicable in the COMPASS kinematic 
regime~\cite{morreale,PSV}.

\section{Recent developments}

Interesting new developments have taken place following the original 
DSSV analysis, mostly related to the advent of new data. We will 
summarize these in the following.

\subsection{Recent DIS and SIDIS data}

\begin{figure}

\vspace*{-8.cm}
\hspace*{4.cm}
\epsfig{figure=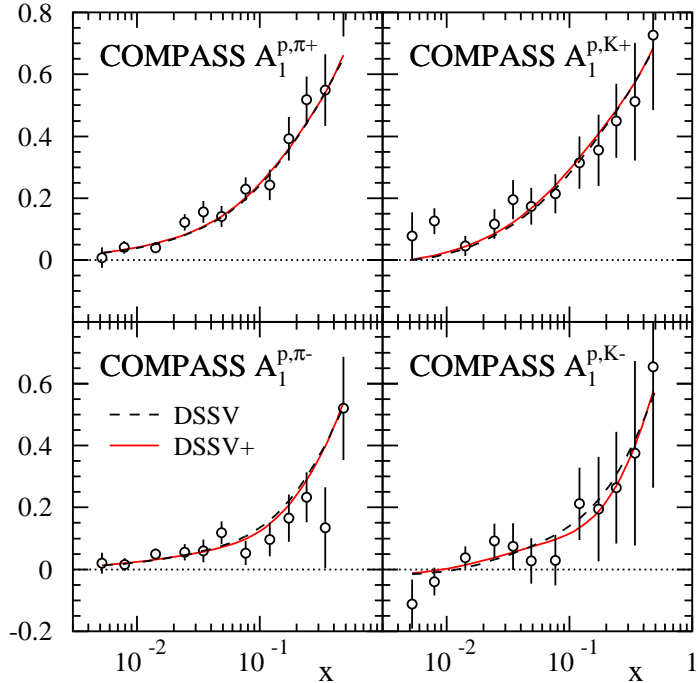,width=0.8\textwidth}

\vspace*{-0.5cm}
\caption{\label{fig:newsidis} COMPASS results~\cite{Alekseev:2010ub} 
for SIDIS spin asymmetries on a proton target, compared to DSSV~\cite{dssv} 
and DSSV+ fits~\cite{msDIS}.}  
\end{figure}
Recently, the COMPASS collaboration has published
new DIS~\cite{Alekseev:2010hc} and SIDIS~\cite{Alekseev:2009ci,Alekseev:2010ub} data. 
The latter extend the coverage in $x$ down to about $x\simeq 5\times10^{-3}$, almost
an order of magnitude lower than the kinematic reach of the 
HERMES data~\cite{ref:hermes-a1he3-sidisn}
used in the DSSV global analysis of 2008~\cite{dssv}.
For the first time, the new results comprise measurements of identified pions and 
kaons taken with a longitudinally polarized proton target.
Clearly, these data can have a significant impact on fits of helicity PDFs 
and estimates of their uncertainties.
In particular, the new kaon data are expected to serve as an important check of the validity of the 
strangeness density obtained in the DSSV analysis discussed above, 
which instead of favoring a negative 
polarization as in most fits based exclusively on DIS data, prefers a vanishing or perhaps 
even slightly positive $\Delta s$ in the measured range of $x$.

Figure~\ref{fig:newsidis} shows a detailed comparison~\cite{msDIS} 
between the new proton SIDIS spin
asymmetries from COMPASS~\cite{Alekseev:2009ci,Alekseev:2010ub} and
the original DSSV fit (dashed lines). Also shown is the result of a re-analysis at 
NLO~\cite{msDIS} accuracy (denoted as ``DSSV+'') based on the updated data set.
The differences between the DSSV and the DSSV+ fits are hard to notice,
both for identified pions and kaons. 
The total $\chi^2$ of the fit drops only by a few units upon refitting, 
which is not really a significant improvement for a PDF analysis
in view of non-Gaussian theoretical uncertainties. The change in $\chi^2$ is also
well within the maximum $\Delta \chi^2/\chi^2=2\%$ criterion adopted
in the original DSSV global analysis \cite{dssv}. Overall, upon refitting, one
finds a trend towards smaller net polarization 
for $\Delta \bar{u}$ and $\Delta \bar{d}$ in the range $0.001\le x \le 1$
than in DSSV.

\begin{figure}

\vspace*{-0.5cm}
\hspace*{2cm}
\epsfig{figure=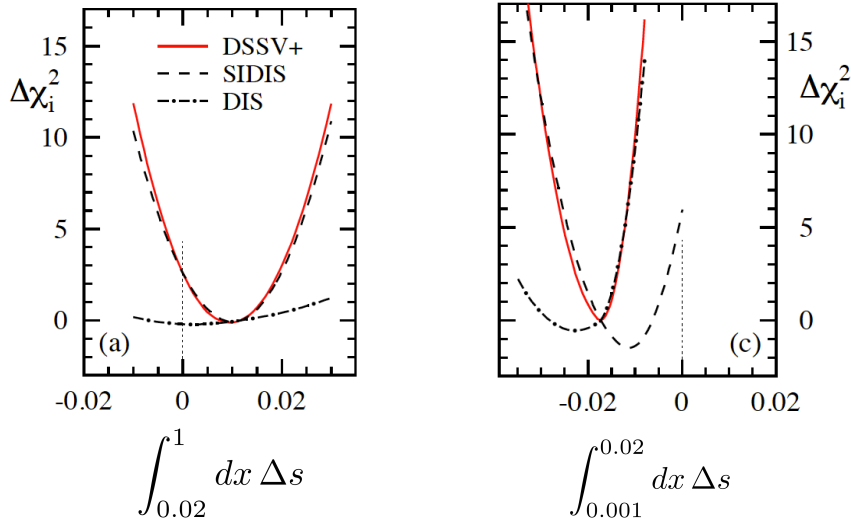,width=\textwidth}

\vspace*{-5.5cm}
\caption{\label{fig:profiles-s}  $\chi^2$ profiles for the truncated first moment
  of $\Delta s$ in two different $x$ intervals, $0.02\le x \le 1$ (left) and
$0.001\le x \le 0.02$ (right).}
  \end{figure}

As we saw earlier,  the original DSSV fit~\cite{dssv} found an interesting feature 
for the strangeness helicity distribution: $\Delta s$ was found to be small
and slightly positive at medium-to-large $x$, but has a significantly {\it negative} 
first moment in accordance with expectations based on SU(3) symmetry and fits to DIS data only.
To investigate this issue further, we present in Fig.~\ref{fig:profiles-s}
the $\chi^2$ profiles for two different intervals in $x$: 
$0.02\le x \le 1$ (left) and $0.001\le x \le 0.02$ (right). 
The profiles in Fig.~\ref{fig:profiles-s} clearly show that the result for $\Delta s$ for 
$0.001\le x \le 0.02$ is a compromise between DIS and SIDIS data, the latter favoring 
less negative values. Interestingly though, the new COMPASS SIDIS data, which extend 
towards the smallest $x$ values so far, actually show some preference for a slightly 
negative value for $\Delta s$ as well. For $0.02\le x \le 1$ everything is determined by SIDIS data, 
and all sets consistently ask for a small, slightly positive strange quark polarization. 
There is no hint of a tension with DIS data here as they do not provide a useful constraint at 
medium-to-large $x$. We note that at low $x$ most SIDIS sets in the original DSSV fit 
give indifferent results. We also mention that in the range $x>0.001$ the hyperon decay 
constants, the so-called $F$ and $D$ values, do not play a significant role in 
constraining $\Delta s(x)$. To quantify possible SU(3) breaking effects one needs to 
probe $\Delta s(x)$ at even smaller values of $x$, for instance in SIDIS at a future 
EIC \cite{EAC}. We finally note that the HERMES and COMPASS data
are consistent in the region of overlap, $0.02\le x \le 1$.

Clearly, all current extractions of $\Delta s$ from SIDIS data suffer from a 
significant dependence on kaon FFs, see, e.g., 
Refs.~\cite{Alekseev:2009ci,Alekseev:2010ub}, 
and better determinations of $D^K(z)$ are highly desirable. 
Contrary to other fits of FFs \cite{2008afa},
only the DSS sets \cite{DSS} provide a satisfactory
description of pion and kaon multiplicities in the same kinematic range where we have
polarized SIDIS data. 

\subsection{$W$ bosons at RHIC}

We have seen in the previous section that the SIDIS data provide
some insights into the flavor structure of the 
polarized sea distributions of the nucleon, albeit with still fairly
large uncertainties. Complementary and clean information on 
$\Delta u,\,\Delta \bar{u},\, \Delta d$, and $\Delta \bar{d}$ will
come from $pp\to W^{\pm}X$ at RHIC, 
where one exploits the maximally parity-violating couplings of the
produced $W$ bosons to left-handed quarks and right-handed 
anti-quarks~\cite{rhicrev,soffer1}. These give rise to a
single-longitudinal spin asymmetry,
\begin{equation}
A_L\equiv\frac{\sigma^{+}-\sigma^{-}}{\sigma^{+}+\sigma^{-}} \, ,
\label{eq:aldef}
\end{equation}
for the processes $\vec{p}p\to \ell^{\pm} X$, where the arrow denotes
a longitudinally polarized proton and $\ell=e$ or $\mu$ is the
charged decay lepton.
The high scale set by the $W$ boson mass makes it possible to 
extract quark and anti-quark polarizations from inclusive lepton 
single-spin asymmetries in $W$ boson production with minimal theoretical 
uncertainties, as higher order and sub-leading terms in the perturbative 
QCD expansion are suppressed~\cite{kamal,nadolsky,asmita,ddfwv}. 

For $W^-$ production, neglecting all partonic processes but the 
dominant $\bar{u}d \to W^-$ one, the spin-dependent cross section
in the numerator of the asymmetry is found to be proportional to the 
combination
\begin{equation}
\Delta \sigma \, \propto \, \Delta \bar{u}(x_1)\otimes d(x_2)(1-\cos\theta)^2-
\Delta d(x_1)\otimes  \bar{u}(x_2)(1+\cos\theta)^2\, ,
\label{eq:w-lo}
\end{equation} 
where $\theta$ is the polar angle of the electron in the partonic
c.m.s., with $\theta=0$ in the forward direction of
the polarized parton. At large negative pseudorapidity 
$\eta_{\mathrm{lept}}$ of the charged lepton, 
one has $x_2\gg x_1$ and $\theta\gg\pi/2$. In this case, the 
first term in Eq.~(\ref{eq:w-lo}) strongly dominates, since 
the combination of parton distributions, $\Delta\bar{u}(x_1)
d(x_2)$, and the angular factor, $(1-\cos\theta)^2$, each dominate 
over their counterpart in the second term. Since the denominator
of $A_L$ is proportional to $\bar{u}(x_1)\otimes d(x_2)(1-\cos\theta)^2+
d(x_1) \otimes \bar{u}(x_2)(1+\cos\theta)^2$, the asymmetry provides a
clean probe of $\Delta\bar{u}(x_1)/\bar{u}(x_1)$ at medium
values of $x_1$.  By similar 
reasoning, at forward rapidity $\eta_{\mathrm{lept}}\gg 0$ the 
second term in Eq.~(\ref{eq:w-lo}) dominates, giving access
to $-\Delta d(x_1)/d(x_1)$ at relatively high $x_1$. 

For $W^+$ production, one has the following structure of the 
spin-dependent cross section:
\begin{equation}
\Delta \sigma \, \propto \, \Delta \bar{d}(x_1)\otimes u(x_2)(1+\cos\theta)^2-
\Delta u(x_1) \otimes \bar{d}(x_2)(1-\cos\theta)^2\, .
\label{eq:w+lo}
\end{equation} 
Here the distinction of the two contributions by considering 
large negative or positive lepton rapidities is less clear-cut than
in the case of $W^-$. For example, at negative $\eta_{\mathrm{lept}}$
the partonic combination $\bar{d}(x_1)u(x_2)$ will dominate, but
at the same time $\theta\gg\pi/2$ so that the angular 
factor $(1+\cos\theta)^2$ is small. Likewise, at positive 
$\eta_{\mathrm{lept}}$ the dominant partonic combination 
$\Delta u(x_1) \bar{d}(x_2)$ is suppressed by the angular factor.
So both terms in Eq.~(\ref{eq:w+lo}) will compete essentially 
for all $\eta_{\mathrm{lept}}$ of interest. Our global 
analysis technique is of course suited for extracting information on the 
polarized PDFs even if there is no single dominant partonic 
subprocess. The NLO corrections to the single-inclusive lepton
cross sections have recently been presented in~\cite{ddfwv} in
a way tailored to use in the global analysis framework.

\begin{figure}[t]

\vspace*{0.5cm}
\hspace*{5.2cm}
\epsfig{figure=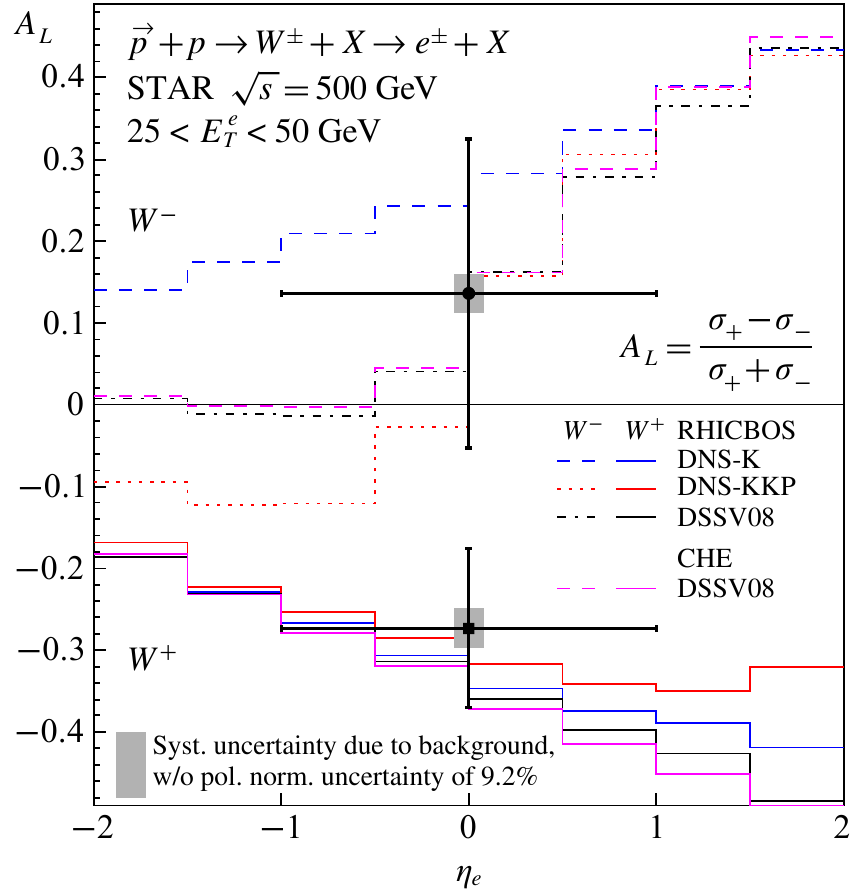,width=0.37\textwidth}

\vspace*{0.4cm}
\hspace*{2.9cm}
\epsfig{figure=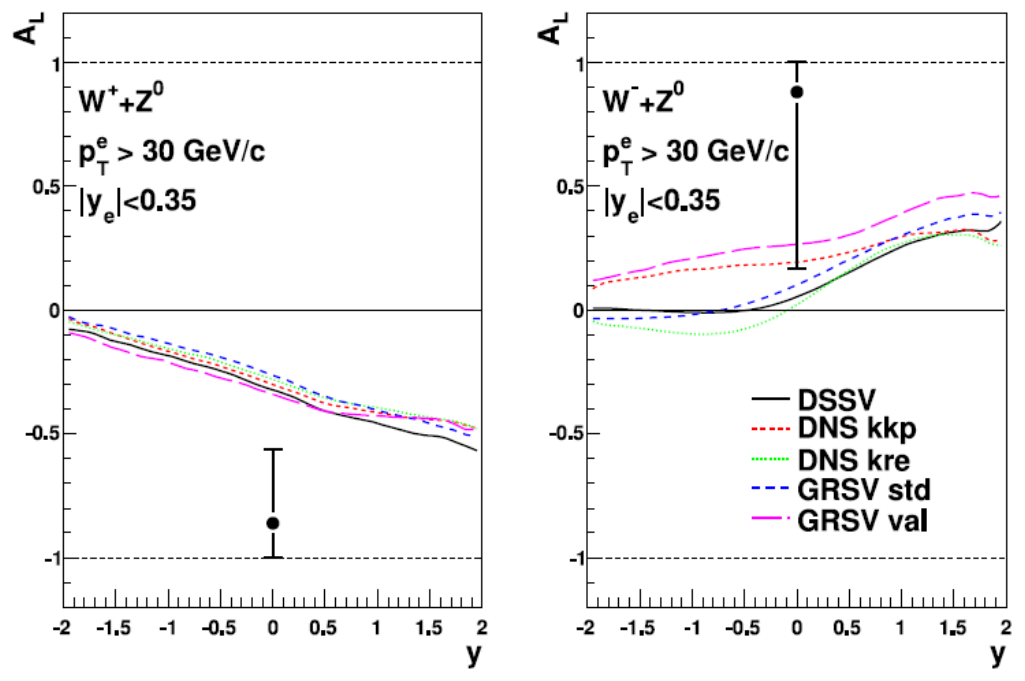,width=0.65\textwidth}

\vspace*{-0.1cm}
\caption{\label{fig:W} Published STAR~\cite{starW} (top) and PHENIX~\cite{phenixW} (bottom)
data for the single-helicity asymmetry $A_L$ in $W^\pm$ production at RHIC.}
\end{figure}

Figure~\ref{fig:W} shows the first published data from RHIC for 
$A_L$ in $W^\pm$ production~\cite{starW,phenixW}.  For now, the 
statistical uncertainties are still large. However, already now a large
negative asymmetry is seen for the case of $W^+$ production, resulting
primarily from the positive up-quark polarization in the proton (see Eq.~(\ref{eq:w+lo})).
Clearly, there is a large potential in future $W$-measurements at RHIC.

\subsection{New constraints on $\Delta g$}

\begin{figure}

\vspace*{-0.5cm}
\hspace*{2cm}
\epsfig{figure=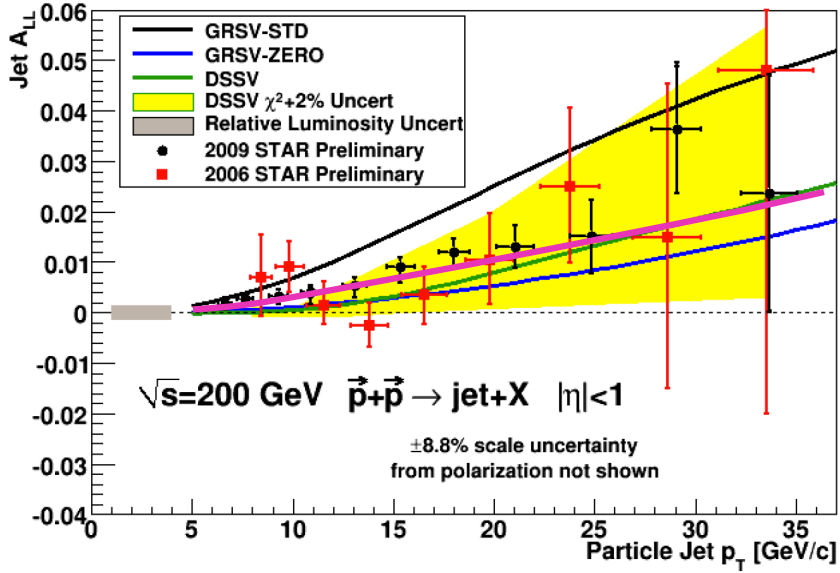,width=0.9\textwidth}

\vspace*{-3.5cm}
\caption{\label{starg} Preliminary STAR run-9 data~\cite{starDG} for the double-helicity
asymmetry $A_{LL}$ for single-inclusive jet production, as a function of jet transverse
momentum for $|\eta|<1$. The preliminary run-6 data are shown as well. The theoretical
curves are as described in the caption. The additional solid magenta line gives
the result for a special DSSV set of polarized parton distributions, for which the truncated 
moment of $\Delta g$ over the region $0.001\leq x\leq 1$ was varied allowing the
total $\chi^2$ of the fit to change by $2\%$.}
\end{figure}

\begin{figure}

\vspace*{-0.5cm}
\hspace*{2cm}
\epsfig{figure=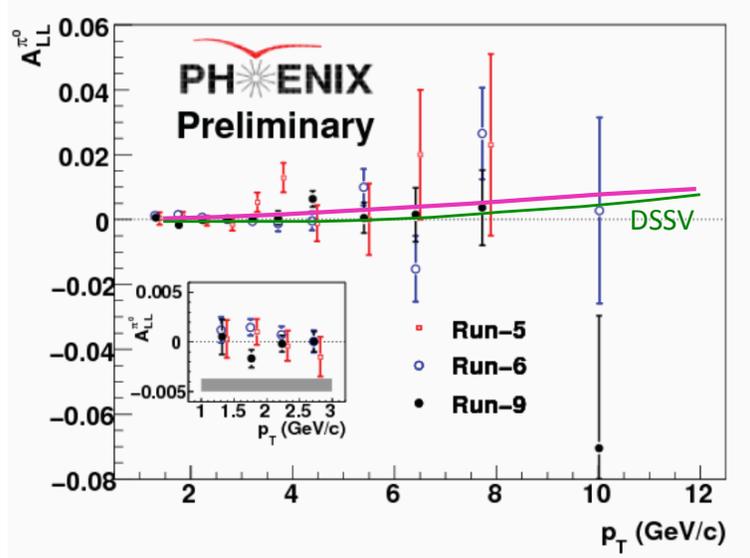,width=0.8\textwidth}

\vspace*{-2cm}
\caption{\label{phenixg} As in Fig.~\ref{starg}, but for the preliminary PHENIX run-9
data~\cite{phenixDG}.}
\end{figure}

The STAR and PHENIX experiments at RHIC have recently presented new preliminary data
from the 2009 run for the double-helicity spin asymmetry $A_{LL}$ for jet and neutral-pion
production, respectively~\cite{starDG,phenixDG}. 
The results are shown in Figs.~\ref{starg} and~\ref{phenixg}.
One can see that the experimental uncertainties are very significantly reduced as compared
to those in the previous run-6 data sets~\cite{starrun6,phenixrun6}. 
An interesting feature of the new preliminary STAR data is that they
lie consistently above the result for the best-fit DSSV distribution for 
jet transverse momenta below $25$~GeV or so. They do remain well below 
the old GRSV-``standard'' result of~\cite{grsv}, on the other hand. This suggests 
that the spin-dependent gluon distribution may be somewhat different from
zero in the $x$-range where it is constrained by the RHIC data. The thick (magenta)
solid line in the figure shows $A_{LL}$ obtained for a special set of parton
distributions within the DSSV analysis. For this set the truncated moment
of $\Delta g$ over the region $0.001\leq x\leq 1$ was varied, allowing the
total $\chi^2$ to change by $2\%$. Evidently, this set of parton distributions
describes the STAR data rather well. The truncated moment of $\Delta g$
in this set over the $x$-range accessed at RHIC is
\be
\int_{0.05}^{0.2} dx \,\Delta g(x,Q^2=10~\mathrm{GeV}^2)=0.13\;,
\ee
which is just within the range 
\be
\int_{0.05}^{0.2} dx \,\Delta g(x,Q^2=10~\mathrm{GeV}^2)=0.005^{+0.129}_{-0.164}\;
\ee
quoted as more conservative uncertainty ($\Delta \chi^2/\chi^2=2\%$) 
in~\cite{dssv}. Despite the fact
that this really is only an illustration that cannot replace a proper 
re-analysis of the data, it does appears that, for the first time, 
there are indications of non-vanishing gluon polarization in the nucleon. 
Figure~\ref{phenixg} shows the comparison to the new preliminary 
PHENIX $\pi^0$ data~\cite{phenixDG}.
Here the values of $A_{LL}$ are much smaller, which is mostly due
to the fact that lower values of $x$ are probed at the transverse momenta
relevant in the PHENIX measurements. One can see that the data are well described by
both the DSSV set and the special set of polarized parton distributions 
used in Fig.~\ref{starg}.

\section{The future: Electron Ion Collider (EIC)}

\begin{figure}[h]

\vspace*{-10.2cm}
\hspace*{4.5cm}
\epsfig{figure=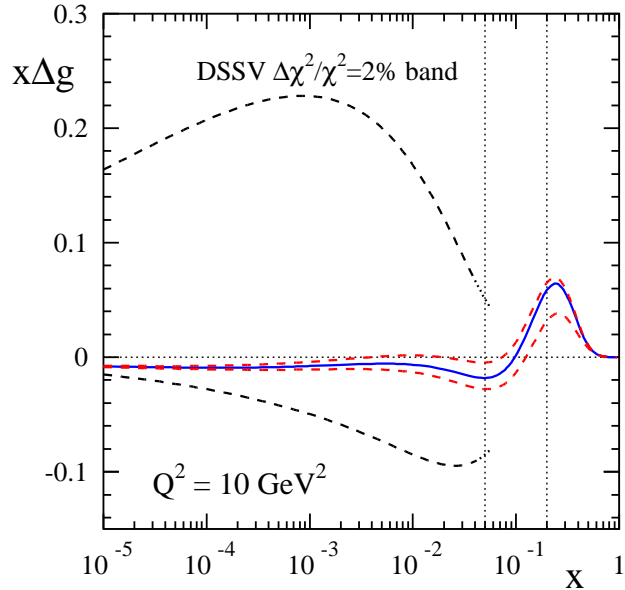,width=0.8\textwidth}

\vspace*{-0.2cm}
\caption{\label{eic} Uncertainty bands for $x\Delta g$ 
referring to $\Delta \chi^2/\chi^2=2\%$ with and without including the generated 
EIC pseudo-data in the fit. The dashed lines correspond to the yellow band for
$\Delta g$ shown in Fig.~\ref{pdfstatus}.}
\end{figure}

An Electron Ion Collider is currently being
considered in the U.S. as a new frontier facility to explore strong-interaction 
phenomena~\cite{Boer:2011fh}. One of its key features would be the 
availability of high-energy, high-luminosity polarized $ep$ collisions 
to probe nucleon spin structure. This would also allow precision 
extractions of $\Delta g$, in particular from scaling violations of 
the proton's spin-dependent structure function $g_1$. Figure~\ref{eic}
shows the results of a recent dedicated phenomenological study~\cite{SaSt}. 
``Pseudo'' EIC-data were generated for collisions of  5 GeV electrons with 
50, 100, 250, and 325 GeV protons and were added to the DSSV global
analysis code. The statistical precision of the data sets for $100-325$ GeV protons
was taken to correspond to about two months of running at the anticipated 
luminosities for eRHIC with an assumed operations efficiency of $50\%$.
For $5\times 50\,\mathrm{GeV}$ an integrated luminosity of $5\,\mathrm{fb}^{-1}$ 
was assumed. The projected uncertainties were used to randomize the pseudo-data by 
one sigma around their central values determined by the DSSV set of PDFs.
With these projected EIC data with their estimated uncertainties, a re-fit of the DSSV 
polarized parton distribution functions was performed. The results are
shown in Fig.~\ref{eic}. As one can see, with EIC data it should be possible to 
map the currently completely undetermined shape of $\Delta g$ for
$10^{-4}\lesssim x\lesssim 0.01$ to an accuracy of about $\pm 10\%$ or better. 

\section*{Acknowledgments}
This work was supported in part by the U.S.\ Department of Energy 
(contract number DE-AC02-98CH10886), and by 
CONICET, ANPCyT and UBACyT.


\end{document}